\documentclass[preprint,showpacs,preprintnumbers,amsmath,amssymb]{revtex4}
\usepackage{graphicx}
\usepackage{dcolumn}
\usepackage{bm}

\begin{document}

\title{Tachyon Field in Intermediate Inflation}

\author{Sergio del Campo}
 \email{sdelcamp@ucv.cl}
\affiliation{ Instituto de F\'{\i}sica, Pontificia Universidad
Cat\'{o}lica de Valpara\'{\i}so, Casilla 4059, Valpara\'{\i}so,
Chile.}
\author{Ram\'on Herrera}
\email{ramon.herrera@ucv.cl} \affiliation{ Instituto de
F\'{\i}sica, Pontificia Universidad Cat\'{o}lica de
Valpara\'{\i}so, Casilla 4059, Valpara\'{\i}so, Chile.}
\author{Adolfo Toloza}
\email{toloza@ucv.cl} \affiliation{ Instituto de F\'{\i}sica,
Pontificia Universidad Cat\'{o}lica de Valpara\'{\i}so, Casilla
4059, Valpara\'{\i}so, Chile.}
\date{\today}

\begin{abstract}
The tachyonic  inflationary universe model in the context of
intermediate inflation   is studied. General
 conditions  for this model to be realizable are
 discussed. In the slow-roll approximation,  we
 describe in great details the characteristics of this model.
\end{abstract}


\maketitle

\section {Introduction}

Nowadays cosmology presents  explosive activity which is
principally due to  theoretical developments and  accurate
astronomical data. In this context, cosmology allows to use
astrophysics to perform tests of fundamental theories, otherwise
inaccessible to terrestrial accelerators. In order to do this task
it is necessary to perform a study about how the Universe evolves
during its different periods. In fact, this study leads to
considering at some stage in the early Universe an inflationary
phase, which is to date the most compelling solution to many
long-standing problems of the big bang model (horizon, flatness,
monopoles, etc.) \cite{guth,infla}.

The source of inflation is a scalar field (the inflaton field)
which plays an important role in providing  a causal
interpretation of the origin of the observed anisotropy of the
cosmic microwave background (CMB) radiation, and also the
distribution of large scale structures \cite{astro,astro2}. The
identity of this scalar field may be found by considering one of
the extensions of the standard model of particle physics based on
grand unified theories, supergravity, or string theory.

In what  concerns  the scalar inflaton field, its dynamics usually
is determined by the Klein-Gordon action. However, more recently,
and motivated by string theory, it is extremely natural to
consider other nonstandard scalar field action.  In this context,
the deep interplay between small-scale nonperturbative string
theory and large-scale braneworld scenarios has aroused interest
in a tachyon field as an inflationary mechanism, especially in the
Dirac-Born-Infeld action formulation as a description of the
D-brane action\cite{Se01,Se02,Ga,BedRdWEy,Kl,KuNi}. Here, rolling
tachyon matter is associated with unstable D-branes. The decay of
these D-branes produces a pressureless gas with finite energy
density that resembles classical dust. Cosmological implications
of this rolling tachyon were first studied by Gibbons\cite{Gi},
and in this context it is quite natural to consider scenarios in
which inflation is driven by the rolling tachyon field. In recent
years the possibility of an inflationary phase described by the
potential of a tachyon field has been considered in a quite large
diversity of
topics\cite{FaTy,ChGhJaPa,KoLi,LiHaLi,LePe,NoOd,KaKaLiMaMcTr,GuZh,LiLi,AgLa01,AgLa02,Ca,MeWa,Ho01,Ho02,BadCHeLaSa,BadCHeLa}.

On the other hand, string/M-theory  suggests that in order to have
a ghost-free action high order curvature, invariant corrections to
the Einstein-Hilbert action must be proportional to the
Gauss-Bonnet (GB) term{\cite{BD}}. GB terms arise naturally as the
leading order of the $\alpha$ expansion to the low-energy string
effective action, where $\alpha$ is the inverse string
tension{\cite{KM}}. This kind of theory has been applied to
possible resolution of the initial singularity
problem{\cite{ART}}, to the study of black-hole solutions{\cite{
Varios1}}, and  accelerated cosmological solutions{\cite{
Varios2}}. In particular, very recently, it has been found that
for a dark energy model the GB interaction in four dimensions with
a dynamical dilatonic scalar field coupling leads to a solution of
the form $a(t) = a_0 \exp{(A t^{f})}${\cite{Sanyal}}.   Here, the
constant $A$ is given by $A= \frac{2}{\kappa n}$ and
$f=\frac{1}{2}$, with $\kappa^2 = 8 \pi G$, and $n$ is a constant.
Therefore, we may argument  that intermediate inflation comes from
an effective theory at a low dimension of a more fundamental
string theory.

In general,  in the context of inflation we have the particular
scenario of ''intermediate inflation'', in which  the scale factor
evolves as $a(t)=\exp(A t^f)$. Therefore, the expansion of the
Universe is slower than standard de- Sitter inflation
($a(t)=\exp(H t)$), but faster than power law inflation ($a(t)=
t^p; p>1$). The intermediate inflationary model was introduced as
an exact solution for a particular scalar field potential of the
type $V(\phi)\propto \phi^{-4(f^{-1}-1)}$, where $f$ is a free
parameter\cite{Barrow1}. With this sort of potential, and with
$1>f> 0$, it is possible in the slow-roll approximation to have a
spectrum of density perturbations, which presents a
scale-invariant spectral index $n_s=1$, i.e., the so-called
Harrizon-Zel'dovich spectrum of density perturbations, provided
$f$ takes the value of 2/3\cite{Barrow2}. Even though  this kind
of spectrum is disfavored by the current Wilkinson Microwave
Anisotropy Probe (WMAP) data\cite{astro,astro2}, the inclusion of
tensor perturbations, which could be present at some point by
inflation and parametrized by the tensor-to-scalar ratio $r$, the
conclusion that $n_s \geq 1$ is allowed providing  that the value
of $r$ is significantly nonzero\cite{ratio r}. In fact, in Ref.
\cite{Barrow3} it was shown that the combination $n_s=1$, and
$r>0$ is given by a version of the intermediate inflation  in
which the scale factor varies as $a(t)\propto e^{t^{2/3}}$ within
the slow-roll approximation.

In this paper we would like to study intermediate inflationary
Universe model in which a tachyon field theory is taken into
account. We will solve the Friedmann and tachyon field equations
for an intermediate expansion of the scale factor and results will
be compared with those obtained in the same situation, but where a
standard scalar field is considered. We should note that this sort
of problem has been studied in the literature\cite{AF}. Here, in
this paper we would like to go further and thus constraint the
parameters of our model by taking into account the WMAP 3 and 5 yr
data.

The outline of the paper is as follows: The next section presents
a short review of the tachyonic-intermediate inflationary phase.
Section \ref{sectp} deals with the calculations of cosmological
perturbations in general term. Finally, in Sect.\ref{conclu} we
conclude with our finding.


\section{Tachyon-Intermediate Inflation Model}

We begin by writing the Friedmann equation for a flat universe and
the conservation equation,
\begin{eqnarray}
\label{friedmann}H^2&=&\frac{\kappa^2}{3} \rho,
\end{eqnarray}
and
\begin{eqnarray}
\label{conservation}\dot{\rho}+3 H (\rho + p)&=&0,
\end{eqnarray}
where  $H=\dot{a}/a$ denotes the Hubble parameter, $\kappa^2=8 \pi
G=8 \pi/ m_p^2$ ($m_p$ represents the Planck mass) and the dots
mean derivatives with respect to the cosmological time $t$. For
convenience we will use units in which $c=\hbar=1$.  For a
tachyonic  field the energy density and the pressure are given by
\begin{eqnarray*}
\rho = \frac{V(\phi)}{\sqrt{1-\dot{\phi}^2}}\ \; , \,
\,{\mbox{and}}\;\;\;\,\;\; p=-V(\phi)\sqrt{1-\dot{\phi}^2}\,,
\end{eqnarray*}
respectively. Here $\phi$ is the tachyonic scalar  field and
$V(\phi)$ its scalar potential, which satisfies $dV/d\phi<0$, and
$V(\phi\rightarrow\infty)\rightarrow 0$, characteristic of any
tachyon field potential\cite{Se01}.

From Eqs.(\ref{friedmann}) and (\ref{conservation}) we get for the
velocity of the tachyonic scalar field, and its evolution equation
becomes

\begin{eqnarray}
\label{phi}\dot{\phi}=\sqrt{-\frac{2 \dot{H}}{3 H^2}},
\end{eqnarray}
and
\begin{eqnarray}
\label{V}\frac{\ddot{\phi}}{1-\dot{\phi}^2}+3 H
\dot{\phi}=-\frac{V'}{V},
\end{eqnarray}
respectively. Here, $V'=\partial V(\phi)/\partial \phi$.

On the other hand, in intermediate inflation it is assumed that
the scale factor follows the law
\begin{eqnarray}
\label{scalefactor}a(t)=a_0\exp{(A t^f)}\ \ ; \ \ 0<f<1 ,
\end{eqnarray}
where $A>0$ has the dimension of $m_p^f$. Note that this
assumption immediately determines the behavior of $\dot{\phi}$ and
$V'/V$, as we can see from Eqs.(\ref{phi}) and (\ref{V}). Note
also that $\dot{H}<0$, since $0<f<1$. From  Eqs.(\ref{phi}) and
(\ref{V}) we get for the scalar field, $\phi$, and the scalar
potential $V(\phi)$

\begin{eqnarray}
\label{phiint}\phi&=&\phi_0+\bigg[\frac{8 (1-f)}{3 A f (2-f)^2
}\bigg]^\frac{1}{2}t^{\frac{2-f}{2}},
\end{eqnarray}

and

\begin{eqnarray}
\label{potential}V(\phi)=\alpha\,(\phi-\phi_0)^{-4(1-f)/(2-f)}\;\sqrt{1-B\,(\phi-\phi_0)^{-2f/(2-f)}},
\end{eqnarray}
with
$$
\alpha=\frac{3}{\kappa^2}\,\left[
A\,f\,\left(\frac{3\,(2-f)^2}{8(1-f)}\right)^{(f-1)}\right]^{2/(2-f)},
$$
and
$$
B=2\,\left[\frac{(1-f)}{3\,A\,f}\right]^{2/(2-f)}\,\left[\frac{(2-f)^2}{8}\right]^{-f/(2-f)},
$$
 respectively.

The Hubble parameter as a function of  $\phi$ becomes
\begin{equation}
H(\phi)=\sqrt{\frac{\alpha\,\kappa^2}{3}}\;(\phi-\phi_0)^{2(f-1)/(2-f)}.\label{HH}
\end{equation}

 Without loss of generality $\phi_0$ can be
taken to be vanished.


During the inflationary epoch the energy density associated to the
 tachyon field is of the order of the potential, i.e.
$\rho\sim V$. Assuming   the set of slow-roll conditions, i.e.
$\dot{\phi}^2 \ll 1$ and $\ddot{\phi}\ll 3H\dot{\phi}$
\cite{Gi,FaTy}, Eqs.(\ref{friedmann}) and (\ref{V}) become
\begin{equation}
\label{slfried}H^2\approx \frac{\kappa^2}{3} V,
\end{equation}
 and
\begin{equation}
\label{slcont}\frac{V'}{V}\approx-3 H \dot{\phi}.
\end{equation}
In this approximation the scalar field potential, $V(\phi)$
becomes

\begin{eqnarray*}
V(\phi)&\approx&\alpha \phi^{-2 \beta},
\end{eqnarray*}
where
\begin{eqnarray*}
\beta\equiv \frac{2(1-f)}{2-f}.
\end{eqnarray*}
Note that this result is also obtained from Eq. (\ref{potential})
by taking  $1\,\gg B\,\phi^{-2 f/(2-f)}$.

Note that this kind of potential does not present a minimum. This
characteristic of the potential makes the study of reheating of
the Universe in a  nonstandard way\cite{SRc}.

At this stage, it is convenient to introduce the slow-roll
parameters $\varepsilon$ and $\eta$, such that
\begin{eqnarray}
\label{epsilon}\varepsilon =-\frac{\dot{H}}{H^2}\approx
\frac{V'^2}{\kappa^2 V^3}\approx \frac{4 \beta^2}{\kappa^2 \alpha}
\phi^{2(\beta-1)},
\end{eqnarray}
and
\begin{eqnarray}
\label{eta}\eta=-\frac{\ddot{\phi}}{H\, \dot{\phi}} \approx
\frac{V'^2}{\kappa^2 V^3}-\frac{V''}{\kappa^2 V^2}\approx -\frac{2
\beta}{\kappa^2 \alpha} \phi^{2(\beta-1)},
\end{eqnarray}
which will be useful  in the study of  perturbations of the model.


On the hand, the number of e-folds between  two different values
$\phi(t=t_1)=\phi_1$ and $\phi(t=t_2)=\phi_2>\phi_1$  is given by

\begin{eqnarray}
N=\int_{t_1}^{t_2} H dt = \frac{\kappa^2 \alpha}{4 \beta(1-\beta)}
\Big[\phi_2^{-2(\beta-1)}-\phi_1^{-2(\beta-1)}\Big].
\end{eqnarray}
Here, we have used  Eq.(\ref{phiint}). This expression allows us
to determine the value of $\phi_2$ in term of $N$, $A$, and $f$.

Following Refs.\cite{Barrow1,Barrow2}, $\phi_1$ it is obtained
from the condition $\varepsilon=1$ (at the beginning of
inflation), that is, at $\phi_1^{2(\beta-1)}=\frac{\kappa^2
\alpha}{4 \beta^2}$.



\section{Perturbation}\label{sectp}
In this section we will study the scalar and tensor perturbations
for our model. The  general expression for the perturbed metric of
the flat Friedmann-Robertson-Walker is
\begin{eqnarray*}
ds^2&=&-(1+2 B) dt^2+ 2 a(t) D_{,\,i}dx^i dt + a^2(t) [(1-2 \psi)
\delta_{ij}+2 E_{,i,j}+2 h_{ij}] dx^i dx^j,
\end{eqnarray*}
where  $B$, $D$, $\psi$, and $E$ are the scalar-type metric
perturbations, and $h_{ij}$ characterizes the transverse-traceless
tensor perturbation. For a tachyon field in the slow-roll
approximation the power spectrum of the curvature perturbation
becomes \cite{Hwang:2002fp}

\begin{eqnarray}
\label{scalar}\mathcal {P}_\mathcal{R}= \Big(\frac{H^2}{2 \pi
\dot{\phi}}\Big)^2\frac{1}{Z_S}  \approx \Big(\frac{H^2}{2 \pi
\dot{\phi}}\Big)^2\frac{1}{V} \approx \frac{\kappa ^6}{12 \pi^2}
\frac{V^4}{V'^2},
\end{eqnarray}
where $Z_S=V\,(1-\dot{\phi}^2)^{-3/2}\approx\,V$ \cite{P}. From
this equation we can derive the spectral index given as
$n_s-1=\frac{d\,\ln \mathcal {P}_\mathcal{R}}{d\, \ln k}$, where
the interval of wave number $k$ is related to the number of
$e$-folds by $d \ln k\simeq d N$. In terms of the slow-roll
parameters it is given in first-order approximation by \cite{FaTy}
\begin{equation}
n_s  \approx\, 1\,-2(\varepsilon+\eta),\label{ns1}
\end{equation}
and from Eqs.(\ref{epsilon}) and (\ref{eta}) we get
\begin{eqnarray*}
n_s &\approx& 1-\frac{4}{\alpha
\kappa^2}\beta(2\beta-1)\phi^{2(\beta-1)}.
\end{eqnarray*}
Since $1>f>0$, we clearly see that the Harrison-Zel'dovich model,
i.e., $n_s=1$ occurs for $\beta=1/2$ or equivalently $f=2/3$. For
$n_s>1$ we have $\beta<1/2$, and $n_s<1$ is for $\beta>1/2$
(recall that $\beta=2(1-f)/(2-f))$.

One of the interesting features of the 5 yr data set from WMAP is
that it hints at a significant running in the scalar spectral
index $dn_s/d\ln k=n_{run}$ \cite{astro,astro2}. From
Eq.(\ref{ns1}) we get  that the running of the scalar spectral
index becomes

\begin{equation}
n_{run}=\left(\frac{4\,V}{V'}\right)\,[\varepsilon_{,\,\phi}+\eta_{,\,\phi}]
\;\varepsilon.\label{dnsdk}
\end{equation}
In models with only scalar fluctuations the marginalized value for
the derivative of the spectral index is approximately $-0.03$ from
WMAP 5 yr data only \cite{astro}.

On the other hand, the generation of tensor perturbations during
inflation would produce  gravitational waves, and its amplitudes
are given by \cite{Bar}
\begin{equation}
{\cal{P}}_\mathcal{T}=8\kappa^2\,\left(\frac{H}{2\pi}\right)^2
\simeq\frac{2\,\kappa^4}{3\pi^2}\,V,\label{ag1}
\end{equation}
where the spectral index $n_g$ is given by $
n_g=\frac{d{\cal{P}}_\mathcal{T}}{d\,\ln k}=-2\,\varepsilon$.

From Eqs.(\ref{scalar}) and (\ref{ag1}) we write  the
tensor-scalar ratio as

\begin{equation}
r=\Big(\frac{\mathcal{P}_\mathcal{T}}{\mathcal{P}_\mathcal{R}}\Big)\approx
\frac{8 V'^2}{\kappa^2 V^3}.\label{ag}
\end{equation}

From expressions (\ref{ns1}) and (\ref{ag}) we write a relation
between $n_s$ and $r$ as

\begin{eqnarray}
\label{standard21}n_s &\approx& 1-\frac{2-3f}{16(1-f)} r,
\end{eqnarray}
i.e., $n_s$ depends linearly with respect to $r$.

Note that Eq.(\ref{standard21}) exactly coincides with the
expression obtained in Ref.\cite{Barrow3}, where  a standard
scalar field was considered. Therefore, it may come  as a surprise
that, on the basic of the intermediate inflation, the trajectories
in the $n_s-r$ plane between standard field and tachyon field can
not be
 distinguished  at lowest order. Actually, this coincidence has
 already been noted in Ref.\cite{Steer:2003yu}.
However, tachyon inflation leads to a deviation at  second order
in the consistency relations, i.e., $n_s=n_s(r)$ . From the same
reference,  the scalar spectral index $n_s$ up to second order in
the slow-roll parameter becomes
\begin{eqnarray}
\label{standard2}n_s &\approx& 1-
\,2(\varepsilon+\eta)-[(2\varepsilon^2+2(2C+3-2\tilde{\alpha})\varepsilon\eta+2C\eta\gamma],
\end{eqnarray}
where $C\simeq -0.72$ is a numerical constant and
$\eta\gamma=(9m_p^4/2)(2V''\,V'/V^4-10V''V'^2/V^5+9V'^4/V^6)$. In
the standard case we have $\tilde{\alpha}=0$,  and
$\tilde{\alpha}=1/6$ for tachyon inflation. Also, at second order,
the  expression for the ratio $r$ is given by $r=16\varepsilon
(1+2C\eta-2\tilde{\alpha}\varepsilon$). These calculations show
that the difference at second order of the consistency relations,
become $n_s^{T}-n_s^{S}\simeq\varepsilon\eta/3 $, where $n_s^{T}$
is the spectral index $n_s$ associated to the tachyon field,
meanwhile $n_s^{S}$ is the same parameter for the standard scalar
field. At this point, we should notice that the relation between
the $r$  and the $n_g$ parameters becomes given by $r = -8 c_s
n_g$ for a tachyonic field, where the speed of sound $c_s$ results
to be $c_s^2=1-(\dot{\phi})^2$\cite{Garriga:1999vw} . However, at
first order it becomes $r\simeq -8 n_g$ \cite{Steer:2003yu}. From
now on, we will consider first-order approximation only, so that
we will work with this latter consistency relation.

In the following we will study the case in which
$\varepsilon\ll\eta$ \cite{FaTy,Steer:2003yu}. In this case, this
condition gives us a constraint for the  values of $f$. To see
this we write down the ratio between $\eta$ and $\varepsilon$ and
we find for the absolute value
\begin{eqnarray*}
\bigg|\frac{\eta}{\varepsilon}\bigg|\approx 1+\frac{3
f-2}{4(1-f)},
\end{eqnarray*}
so, for $\eta>\varepsilon$ we need to have $f>\frac{2}{3}$.

The scalar spectral index $n_s$,  for $\varepsilon\ll\eta$, is
given by
\begin{eqnarray} \label{n_s}n_s &\approx& 1-2\eta.
\end{eqnarray}
From Eq.(\ref{eta}) this expression  is equivalent to
\begin{eqnarray}
\label{ns4} n_s &\approx& 1+\frac{4 \beta}{\kappa^2 \alpha}
\phi^{2(\beta-1)}.
\end{eqnarray}

Using that $\phi_1$ it is obtained from the condition
$\mid\eta\mid=1$ (at the beginning of inflation), then
Eq.(\ref{ns4}) can be re-expressed in terms of the number of
e-folding $N$, resulting in
\begin{eqnarray*}
n_s=1+\frac{2}{1+2\,(1-\beta)\,N}=1-\frac{2(1-f/2)}{(1/2-N)f-1}.
\end{eqnarray*}
Note that,  does  a value does not exist  for $f$ in which
$n_s=1$, in contrast with the standard case \cite{Barrow3} (which
occurs for $f=\frac{2}{3}$). This means that it is not possible to
have a Harrison-Zel'dovich spectrum  in this case.

From Eq.(\ref{n_s}) we can obtain for the running scalar spectral
index
\begin{eqnarray}
\label{nrunn}n_{run}&\approx& \Big(\frac{4 \beta}{\kappa^2
\alpha}\Big)^2 (\beta-1) \phi^{4(\beta-1)}.
\end{eqnarray}

From Eq.(\ref{ns4})  we get a relation between $n_s$ and $n_{run}$
which becomes
\begin{eqnarray*}
n_{s}&\simeq& 1+\sqrt{\frac{2-f}{f}} \ \ \sqrt{-n_{run}}.
\end{eqnarray*}

On the other hand, from Eq.(\ref{ag}) we write the tensor-scalar
ratio as
\begin{eqnarray}
\label{r}r\approx \frac{32}{\kappa^2 \alpha} \beta^2
\phi^{2(\beta-1)},
\end{eqnarray}
and in terms of the  e-folding parameter $N$, we write
\begin{eqnarray}
\label{rN}r=\frac{16 \beta}{1+2(1-\beta)N}.
\end{eqnarray}
Also, from Eqs.(\ref{ns4}) and (\ref{r}) we obtain a relation
between $n_s$ and $r$ which is
\begin{eqnarray}
\label{relation}n_s\approx1+\frac{2-f}{16 (1-f)}\ \ r.
\end{eqnarray}


\begin{figure}[th]
\includegraphics[width=4.0in,angle=0,clip=true]{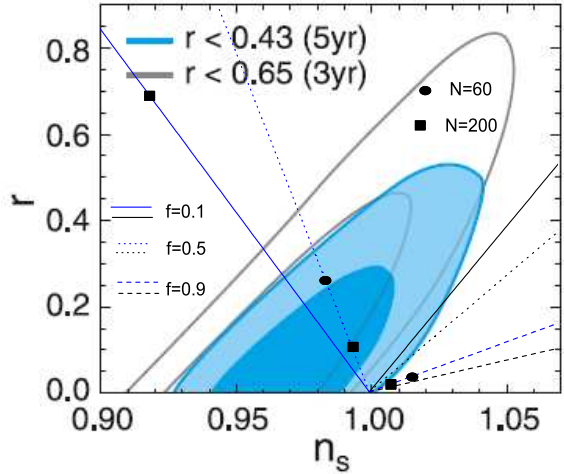}
\caption{  The plot shows $n_s$ versus $r$ for our models and they
are compared with the WMAP data (three year and five year). The
curves in black represent the case $\varepsilon\ll\eta$ and the
blue one specify the tachyon lowest order case for different
values of $f$. The two contours correspond to the $68\%$ and
$95\%$ levels of confidence\cite{astro}. The small black dots and
squares represent the number of e-folds for the values $N=60$ and
$N=200$, respectively.\label{Fig1}}
\end{figure}

In Fig.(\ref{Fig1}) we show the dependence of the tensor-scalar
ratio on the spectral index for different values of the parameter
$f$ for the tachyon lowest order (shown in black line) and the
regimen where $\varepsilon\ll\eta$ (shown by the blue line). In
this plot we have used Eqs.(\ref{standard21}),  and
(\ref{relation}).

The two contour in the plot show the 68$\%$ and 95$\%$ levels of
confidence, for the $r-n_s$ plane, which are defined at $k_0$ =
0.002 Mpc$^{-1}$ \cite{astro}.  The five-year WMAP data places
stronger limits on $r$ (blue) than three-year data
(grey)\cite{Sp}. For tachyon lowest order case any value the
parameter $f$, (restricted to the range $1>f>0$), is well
supported by the data, as can be seen from Fig.(\ref{Fig1}).

On the other hand, when we considered the  regime where
$\varepsilon\ll\eta$ (given in Ref.\cite{FaTy}), we see that for
$f=0.1$ the curve $r=r(n_s)$ barely enters the $95\%$ confidence
region for $r=0.2$,  which corresponds to $N=54$.  From
Fig.(\ref{Fig1})  the best values of $f$ occur for the range
$0.5>f>0$.

\section{Conclusions \label{conclu}}

In this paper we have studied the tachyon-intermediate
inflationary model. We have found in this model an exact solution
of the Friedmann equation for a flat Universe containing a scalar
field, $\phi(t)$, with tachyonic scalar potential, $V(\phi)$. In
the slow-roll approximation we have found a general relation among
the scalar potential and its derivatives. We have also obtained
explicit expressions for the corresponding, power spectrum of the
curvature perturbations $P_{\cal R}$, tensor-scalar ratio $r$,
scalar spectrum index $n_s$, and its running $n_{run}$.  Here, we
noted that Eq.(\ref{standard21}) exactly coincide at lowest order
with the expression obtained in Ref.\cite{Barrow3}, where  a
standard scalar field was studied.

In order to bring some explicit results we have taken the
constraint in the $r-n_s$ plane to first-order in the tachyon
lowest order case and the regime in which $\varepsilon\ll\eta$. In
the tachyon lowest order case, we noted that the parameter $f$,
which lies in the range $1>f>0$, the model   is well supported by
the data as could be seen from Fig.(\ref{Fig1}) for any value of
$f$. But in the other case i.e.  when $\varepsilon\ll\eta$, the
parameter $f$ lies within range $0.5>f>0$, in order to be in
agreement with the current WMAP astronomical data.

We should mention that we have not addressed the phenomena of
reheating and the possible transition to the standard cosmology
(see e.g., Refs.\cite{SRc,u,yo}). A  calculation for the reheating
temperature in the high energy scenario would give new constraints
on the parameters of the models.  We hope to return to this point
in the near future.


\begin{acknowledgments}
 This work was
supported by COMISION NACIONAL DE CIENCIAS Y TECNOLOGIA through
FONDECYT grants N$^0$ 1070306 (SdC) and N$^0$ 1090613 (RH and
SdC).
\end{acknowledgments}


\end{document}